\documentclass[twocolumn,showpacs,preprintnumbers,amsmath,amssymb,superscriptaddress]{revtex4}

\usepackage{graphicx}
\begin{document}
\title{Where Are the Extra $d$ Electrons 
in Transition-Metal Substituted Fe Pnictides?}

\author{H.~Wadati}
\affiliation{Department of Physics and Astronomy, 
University of British Columbia, 
Vancouver, British Columbia V6T 1Z1, Canada}

\author{I.~Elfimov}
\affiliation{Advanced Materials and 
Process Engineering Laboratory, 
University of British Columbia, Vancouver, 
British Columbia V6T 1Z4, Canada}

\author{G.~A.~Sawatzky}
\email{sawatzky@phas.ubc.ca}
\affiliation{Department of Physics and Astronomy, 
University of British Columbia, 
Vancouver, British Columbia V6T 1Z1, Canada}

\date{\today}
\begin{abstract}
Transition-metal substitution in Fe pnictides 
leading to superconductivity is 
usually interpreted in terms of 
carrier doping to the system. 
We report on a density functional calculation 
of the local substitute electron density and 
demonstrate that substitutions like Co and Ni 
for Fe do not carrier dope 
but rather are isovalent to Fe. 
We find that the extra $d$ electrons for Co and Ni 
are almost totally located within the muffin-tin sphere 
of the substituted site. 
We suggest that Co and Ni act more like random scatterers 
scrambling momentum space and washing out parts of the Fermi surface. 
\end{abstract}
\pacs{74.70.Xa, 71.15.Mb, 74.25.Jb, 74.62.Dh}
\maketitle
The discovery of superconductivity in layered Fe 
pnictides \cite{kamihara} is attracting 
a lot of interest because their transition 
temperatures up to $T_c=56$ K \cite{highest} 
are the highest except for cuprates. 
The $A$Fe$_2$As$_2$ 
($A$: alkali-earth metals such as Ca, Sr, and Ba) 
systems are in a spin-density-wave state 
at low temperatures \cite{BaFe2As2n}. Upon hole doping 
by the chemical substitution of $A^{2+}$ ions 
with potassium ions (K$^+$) superconductivity appears \cite{rotter}. 
This is reminiscent of the cuprates 
where the introduction of charge carriers via doping 
indeed seems to control the phase diagram. Surprising is that 
the substitution of Co or Ni for divalent Fe 
also leads to superconductivity \cite{sefat, pccanfield}, 
which by analogy is also mostly 
ascribed to electron doping 
with the explanation that the extra $d$ electrons 
in Co and Ni are donated to the system. 
One should note this implies that Co and Ni would behave like 
Co$^{3+}$ or Ni$^{4+}$. 
Chemical intuition, on the other hand, advocates 
the isovalent nature of the substitution 
with the extra $d$ electrons 
bound to the higher nuclear charge, 
but then leading to questions regarding the role of 
these substitutions for Fe. 
In this paper we study the local electron density 
distribution in the vicinity of the substitutes, 
Co, Ni, Cu and Zn, using density functional 
(DFT) methods. 

Recent theoretical density 
functional theory studies of the changes 
in the density of states close to the Fermi energy 
of the Co substituted materials \cite{sefat} 
suggest a shift in the density of states qualitatively consistent
with a so-called virtual crystal approach in which  basically the extra
nuclear charge on Co is in effect averaged over the whole crystal. The
net result of the virtual crystal approach 
is to good approximation a rigid shift of the Fermi level. It is 
also important to recognize that the density of states is actually not a
ground state property and therfore perhaps a less reliable quantity to
look at within DFT. 
It is interesting therefore to look at the actual charge density 
distribution in the substituted material with DFT 
methods to see where the extra $d$ electrons 
actually reside. By looking at the charge density distribution 
we are looking at a quantity that is a ground-state property 
and therefore at least 
in principle exact in a DFT framework. 
Recent reports 
of a periodic impurity model in DFT 
studied the spin density distribution and found 
that the excess spin density is indeed 
localized at Co sites 
in Co-substituted BaFe$_2$As$_2$ \cite{codope}. 
In this paper, we report on the results of a detailed DFT 
study of a similar periodic impurity model of various substitutes 
(Co, Ni, Cu, Zn, Ru, Rh and Pd) in BaFe$_2$As$_2$ and FeSe. 
The main reasons for including the FeSe substituted 
system is to demonstrate the insensitivity of our conclusions 
to the details of the chemical composition of the host material 
and secondly to be able to handle a larger supercell and 
lower concentration in FeSe to demonstrate the insensitivity 
to the ``impurity'' concentration 
for concentrations below 12.5\%. 
We report a study of the electron density spatial distribution 
which is a basic property of DFT. We show that the extra $d$ 
electrons are almost totally concentrated within the muffin-tin (MT) 
sphere of the substitute atoms consistent with an isovalent scenario 
rather than a virtual crystal scenario. 

The band-structure calculations were 
performed using the linearized augmented plane wave 
method implemented in the WIEN2K package \cite{wien2k}. 
The exchange and correlation effects 
were treated within the generalized gradient 
approximation \cite{gga}. 
As for BaFe$_2$As$_2$, 
we considered the tetragonal I4/mmm structure, 
where we used lattice constants $a=3.9625\ \mbox{\AA}$ 
and $c=13.0168\ \mbox{\AA}$ as obtained in Ref.~\cite{rotter2}. 
The lattice constants of FeSe were taken from 
Ref.~\cite{FeSe2} as 
$a=3.7734\ \mbox{\AA}$ 
and $c=5.5258\ \mbox{\AA}$. 
The effect of substitution 
was taken into account with the supercells 
of Ba$_8$Fe$_{14}$TM$_2$As$_{16}$ (12.5\%) 
and Fe$_{17}$TMSe$_{18}$ (5.6\%) 
(TM = Co, Ni, Cu, Zn, Ru, Rh or Pd), as shown in Fig.~1. 
We adopt the crystallographic notation 
for the directions (11) and (01) 
in the Fe plane of the pure material. 

\begin{figure}
\begin{center}
\includegraphics[width=8cm]{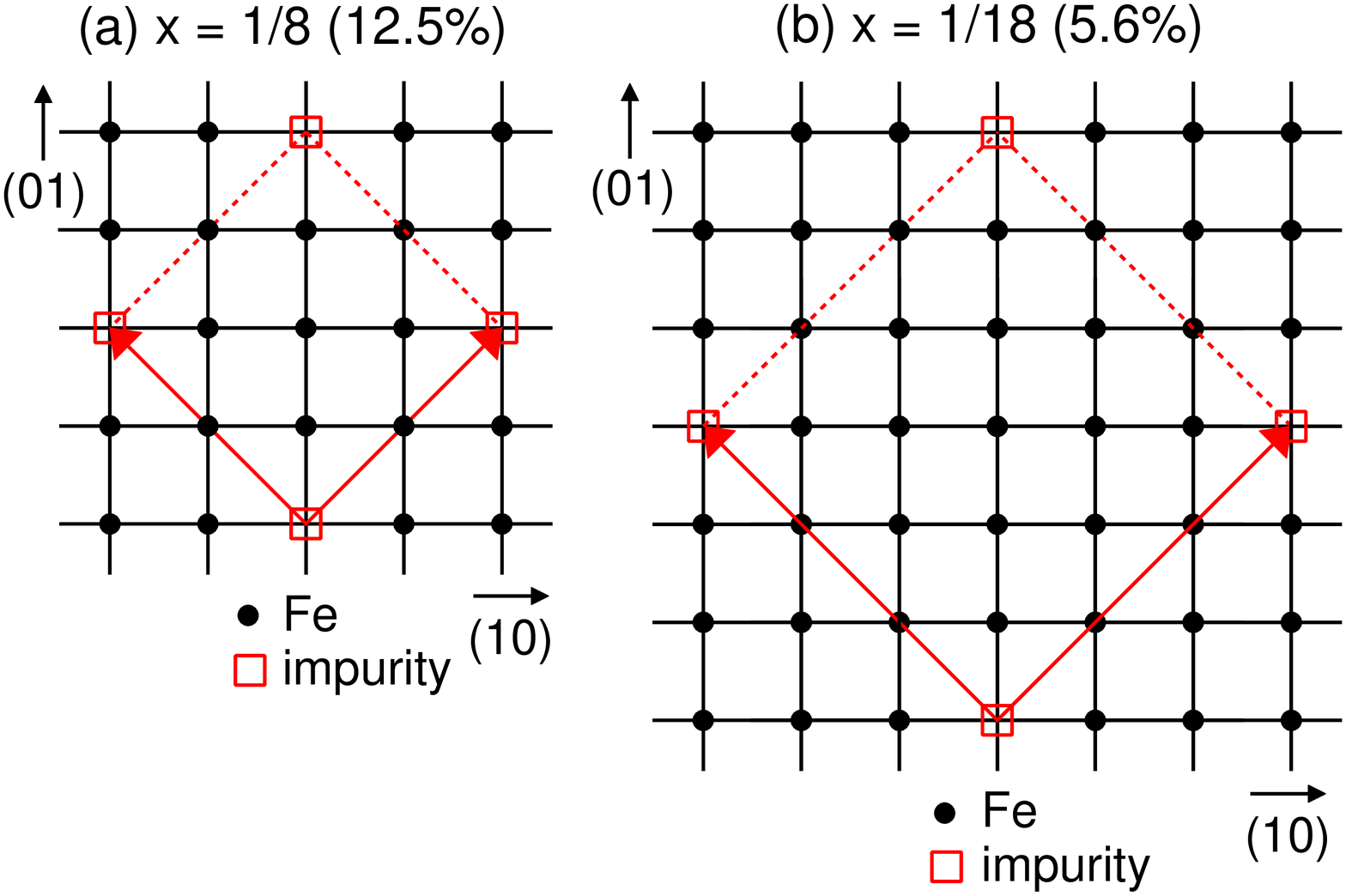}
\caption{(Color online) 
Supercells used for 12.5\% (a) and 5.6\% (b) substitution 
as used in the calculation. 
The arrows define the unit vectors for the supercell. 
The filled circles and open squares denote 
Fe and impurity sites, respectively.}
\label{fig0}
\end{center}
\end{figure}

\begin{figure}
\begin{center}
\includegraphics[width=9cm]{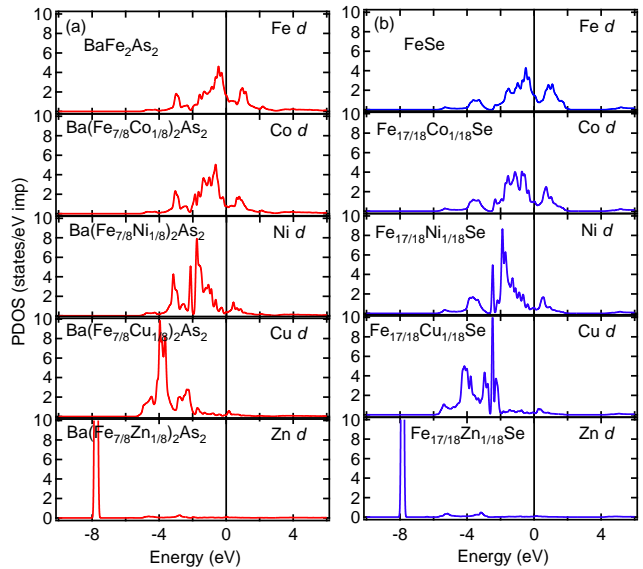}
\caption{(Color online) partial Fe $d$ density of states 
in the pure material and the partial Co, Ni, Cu, 
or Zn $d$ density of states 
in 12.5\% substituted BaFe$_2$As$_2$ (a) and 5.6\% substituted 
FeSe (b). The zero of energy is at the Fermi energy.}
\label{fig1}
\end{center}
\end{figure}

Figure \ref{fig1} shows 
the partial Fe $d$ density of states 
in the pure material and 
the partial Co, Ni, Cu, or Zn $d$ density of states 
in 12.5\% substituted BaFe$_2$As$_2$ (a) and 5.6\% 
substituted FeSe (b). 
The zero of energy is at the Fermi energy. 
The density of states of the substituted materials 
clearly shows a shift relative to the pure material 
but this shift is not uniform. Clearly there 
is a substantial change in the shape itself of the density 
of states indicating a strong influence of 
the changed local potential produced by the substitutes. 
This is most clear in comparing the local projected density of states
for the Ni, Cu and Zn cases with that of the Fe local projected density
of states in the pure material. 
In the extreme and as yet experimentally unstudied case of Zn, 
the Zn $3d$ states form a very narrow "impurity" band 
at $\sim 8$ eV, resembling what is often referred to 
as a shallow core level. The whole evolution of the 
substitute density of states with increasing nuclear 
charge is qualitatively what is expected 
for isovalent substitution. We also note that 
the results for 12.5\% BaFe$_2$As$_2$ and 
5.6\% FeSe 
are almost the same demonstrating that the impurity-impurity 
interaction is already small for the third nearest neighbor 
distances ($\sim 5.6\ \mbox{\AA}$). 
This also demonstrates that the major part of the action 
is confined to the Fe plane being rather 
insensitive to replacing Se with As etc. 

\begin{figure}
\begin{center}
\includegraphics[width=9cm]{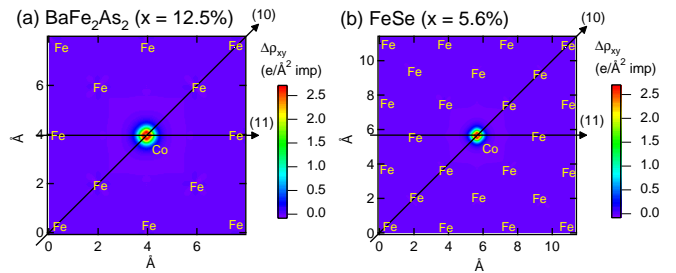}
\caption{(Color online) 
(a) a contour plot of 
the difference in valence electron density integrated 
along $c$-axis direction between 12.5\% Co substituted and 
pure BaFe$_2$As$_2$. 
(b) The same as (a) but for FeSe and 5.6\% substitution.}
\label{fig2}
\end{center}
\end{figure}

\begin{figure}
\begin{center}
\includegraphics[width=9cm]{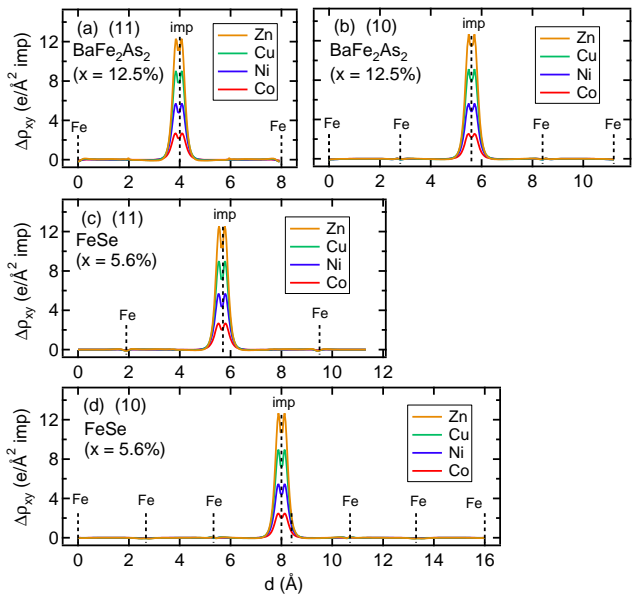}
\caption{(Color online) 
The difference in valence electron density integrated 
along $c$-axis direction between 12.5\% Co substituted and 
pure BaFe$_2$As$_2$ with a cut 
along (11) (a) and (10) (b) direction. 
Panels (c) and (d) are the same as (a) and (b) ,respectively, 
but for FeSe and 5.6\% substitution.}
\label{fig3}
\end{center}
\end{figure}

The electron-density distribution in the ground state 
has a sounder theoretical basis 
in DFT than does the density of states 
since the latter is not a ground-state property, so we 
will focus on the electron density here. 
In Fig.~\ref{fig2} (a), we show the 
valence electron density distribution in the $ab$ 
plane after integration along the $c$-axis direction 
and after subtracting off that of the pure material. 
The valence electron density is defined as the density 
of electrons in the states situated between $-8$ eV 
and the Fermi energy. 
Panel (b) is the same as (a) but 
for FeSe and 5.6\% substitution. 
Both the pure and substituted material were
treated with identical unit cells. 
Aside from the strong increase in the 
electron density close to the substituents, 
there is a quite minor density change in 
the rest of the unit cell. 
A more detailed picture of the change 
in the charge density is shown in
Fig.~\ref{fig3}. Here we plot the change in the 
charge density again integrated
over the c direction with a cut 
along the (11) and (10) directions. 
This figure shows 
the  monotonic increase 
in the local ``impurity'' charge density 
as we go from Fe to Co, Ni, Cu, and Zn. 
One can see that the excess electrons from the impurity 
are concentrated at the impurity site, and there is 
little effect on the electron density distribution 
in the rest of the material. For example, there is 
no observable offset of the electron density 
as would have been expected from a rigid Fermi level shift. 

\begin{figure}
\begin{center}
\includegraphics[width=7cm]{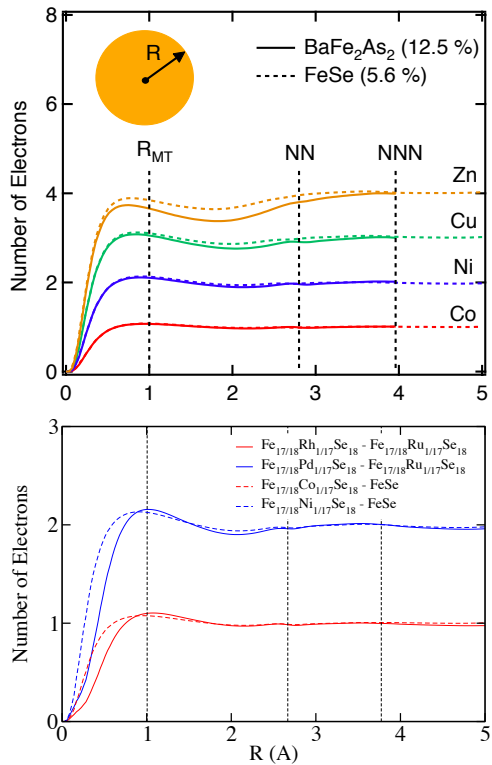}
\caption{(Color online) (a) The integrated change in the 
electron density as a function of the distance $R$ 
from the impurity in 12.5\% 
substituted BaFe$_2$As$_2$ and 5.6\% substituted FeSe. 
As shown in the inset, 
the integration is over the colored disc 
with a radius $R$ and over the $c$ axis. (NN: Nearest neighbor. 
NNN: Next nearest neighbor.) (b) A comparison between $3d$ and
$4d$ substitutions in FeSe. The changes in the electron density 
for the Rh and Pd cases are calculated with respect to 
Ru substitution.}
\label{fig4}
\end{center}
\end{figure}

To determine the number of electrons within the MT 
containing mainly $3d$ electrons in the valence-band region, 
we show in Fig.~\ref{fig4} the integral over the colored disc 
with a radius $R$ and integrated also 
over the $c$ axis (as shown in the inset). 
The result is the radial dependence of the electron density.  
Again here the vertical axis is the 
difference between the pure material with Fe at the origin 
and the substituted material with the substitute at the origin. 
This plot clearly 
demonstrates that for Co one extra electron is located inside the MT 
radius and for Ni it is 2 for Cu it is 3 and for Zn it is 4, 
consistent with isovalent substitution. This 
strongly suggests that these substituents bind locally the 
number of $d$ electrons needed to compensate 
for the change in the nuclear charge. 

To study the distribution of the excess $d$ electrons in case of 
$4d$ substitutes, we calculate the change in the electron density 
of FeSe upon Rh or Pd substitution. There is an issue with 
doing a simple direct difference density with pure Fe because 
the $4d$ orbitals have an extra radial node and are more extended 
than the $3d$ wave functions of Fe and Co or Ni. This causes  a 
relative transfer of charge density from small radii to larger radii 
for the $4d$Õs and because there are 6 or more $4d$ occupied orbitals 
each of which have this contribution it tends to overwhelm the 
contribution from the single extra electron. Rather than compare
these charge densities with pure FeSe, we use FeSe substituted 
with Ru which is isovalent with Fe
has the same number of $d$ electrons and which 
would also have the extended wave functions for all the occupied 
orbitals. Using otherwise the same procedure as above, we get 
the plots shown in Fig.~\ref{fig4}. One can see that Rh is very 
similar to Co and Pd to Ni in that they both have the same number 
of extra electrons i.e. 1 and 2, respectively, as the $3d$ substitutions.
The somewhat larger orbital radius for Rh and Pd 
as compared to Co and Ni is also evident  from the initial slops.

Summarizing, we studied the effects on the electron density 
of various substitutes (Co, Ni, Cu, Zn, Ru, Rh and Pd) 
in BaFe$_2$As$_2$ and FeSe using density functional theory. 
The partial substitute $d$ density of states 
exhibits a strong change compared to that of the projected 
Fe density of states in the pure material 
over a wide energy range. This change is strongly different 
from that obtained by a rigid shift of the Fe projected density 
of states of the pure material. We also presented a DFT calculation 
of the electron density distribution in the unit cell of 
a superlattice of substitutes. This clearly demonstrates 
that the excess $d$ electrons are concentrated within the MT 
sphere at the substitute sites compensating for the increased 
nuclear charge. The charge density in the rest of the unit cell 
is hardly affected. This result strongly suggests 
that Co, Ni, Cu, Zn, Ru, Rh and Pd substitutes for Fe 
should not be considered as ``dopants'' 
because they are isovalent with Fe. 

The question then remains; 
what is the physical origin of the strong influence on $T_c$? 
In answering this, we should not forget that these ``impurities'' 
are located in the Fe plane and therfore will be strong scatterers 
especially for electrons in narrow bands. 
We should also remember that in the real material these 
impurities are randomly distributed and 
so will tend to scramble $k$ space. 
This could have the influence of ``washing out'' 
(parts of) the Fermi surface. 
The parts most strongly effected would be those 
for which the Fermi velocity measured along the band 
dispersion is the smallest. In other words, 
the rather 
flat band contributions to the total Fermi surface 
would be washed out first. 
This could destabilize competing phases based 
on the Fermi surface nesting 
in favour of superconductivity. 

In fact, there are potential smoking gun like experiments that can prove
or disprove 
the importance of impurity scattering in these systems. 
In a recent angle-resolved photoemission spectroscopy paper \cite{LiFeAs}, 
the authors demonstrate very narrow momentum space widths 
for the pure non polar material LiFeAs. 
All other pnictides and FeSe materials exhibit extremely 
wide momentum space widths. Therefore, LiFeAs provides an ideal platform 
for studying momentum space broadening as a function of 
various substitutions. 
Such experiments would be able to distinguish 
accurately between the doping and the impurity scattering effects. 

The authors would like to thank O. K. Andersen 
and A. Damascelli for informative discussions. 
This research was made possible with 
financial support from the Canadian funding organizations 
NSERC, CFI, and CIFAR. 

\bibliography{LVO1tex}

\end{document}